\documentclass[a4, conference]{IEEEtran}
\usepackage{cite}
\usepackage{pgfplots}

\pgfplotsset{compat=1.10}

\usepgfplotslibrary{fillbetween}

\usepackage{bm}
\usepackage{lipsum}
\usepackage{makeidx}
\usepackage{enumerate}
\usepackage{color}
\usepackage{cite}
\usepackage{amsmath,amsthm}   
\usepackage{amssymb}
\usepackage{nomencl}
\usepackage{multirow}
\usepackage{graphicx}
\usepackage{epstopdf}
\usepackage{threeparttable}
\usepackage{multicol}
\DeclareGraphicsExtensions{.pdf,.jpeg,.png,.jpg,.emf,.eps}
\usepackage{calc}
\usepackage{here}
\usepackage{url}

\usepackage{upref}
\usepackage{comment}
\usepackage{times}
\usepackage{dsfont}
\usepackage{epic,eepic}
\usepackage{rawfonts}
\usepackage[T1]{fontenc}
\usepackage{latexsym}
\usepackage{amsfonts}

\usepackage{capitalgreekitalic}

\usepackage{geometry}
\usepackage[font=small,labelfont=bf]{caption}
\usepackage[font=small,labelfont=bf]{subcaption}

\usepackage{xcolor}
\usepackage{tikz,pgfplots}
\usetikzlibrary{calc}
\usetikzlibrary{intersections}
\usetikzlibrary{arrows,shapes}
\usetikzlibrary{positioning}

\theoremstyle{definition}

\allowdisplaybreaks

\begin{document}
\title{
NOMA-based Improper Signaling for MIMO STAR-RIS-assisted Broadcast Channels with Hardware Impairments
}
\author{%
   \IEEEauthorblockN{Mohammad Soleymani$^*$, Ignacio Santamaria$^\dag$, and Eduard Jorswieck$^\ddag$}
   \IEEEauthorblockA{*Signal \& System Theory Group, Universit\"at Paderborn, Germany \\
                     $^\dag$Dept. Communications Engineering, Universidad de Cantabria, Spain\\
$^\ddag$ Institute for Communications Technology, Technische Universit\"at Braunschweig, Germany\\
                     Email: \small{\protect\url{mohammad.soleymani}@sst.upb.de}, \small{\protect\url{i.santamaria@unican.es}}, \small{\protect\url{jorswieck@ifn.ing.tu-bs.de}}
}
}
\maketitle
\begin{abstract}
This paper proposes schemes to improve the spectral efficiency of a multiple-input multiple-output (MIMO) broadcast channel (BC) with I/Q imbalance (IQI) at transceivers by employing  a combination of improper Gaussian signaling (IGS), non-orthogonal multiple access (NOMA) and simultaneously transmit and reflect (STAR) reconfigurable intelligent surface (RIS). When there exists IQI, the output RF signal is a widely linear transformation of the input signal, which may make the output signal improper. To compensate for IQI, we employ IGS, thus generating a transmit improper signal. We show that IGS alongside with NOMA can highly increase the minimum rate of the users. Moreover, we propose schemes for different operational modes of STAR-RIS and show that STAR-RIS can significantly improve the system performance. Additionally, we show that IQI can highly degrade the performance especially if it is overlooked in the design.
\end{abstract} 

\section{Introduction}
A main target of the sixth generation of wireless communication systems (6G) is to improve spectral efficiency (SE) of wireless networks by around 10 times higher than the SE of 5G networks \cite{gong2022holographic}. To this end, we have to encompass some challenges such as interference and hardware impairments (HWI), which may highly affect the system performance. Interference has been always among the main performance limitations of modern wireless communication systems \cite{andrews2014will}. Hence, interference-management techniques such as non-orthogonal multiple access (NOMA) and improper Gaussian signaling (IGS) are expected to continue playing a key role in such networks. Moreover, devices are never ideal in practice, and thus, HWI-aware techniques should be employed to compensate for the non-idealities \cite{soleymani2020rate, soleymani2020improper, soleymani2023rate, javed2019multiple, soleymani2019improper}.

A source for HWI is an imbalance between the in-phase and quadrature signals, which is known as I/Q imbalance (IQI) \cite{soleymani2020improper, javed2019multiple}. In the presence of IQI, the output  transmitted/received signal is a widely linear transformation (WLT) of the input signal, which means that output digital baseband signal can be improper even if the input signal is proper. Note that in a zero-mean improper signal, the real and imaginary parts of the signal are not independent and identically distributed \cite{adali2011complex}. To compensate for IQI, we employ IGS, which is shown to be a powerful interference-management technique to improve the SE and/or energy efficiency of various systems 
\cite{cadambe2010interference, soleymani2019ergodic, soleymani2020improper,  soleymani2019energy, javed2020journey}.

In addition to  interference-management and/or HWI-aware techniques, 6G should employ some emerging technologies like reconfigurable intelligent surface (RIS) to fulfill the ever increasing SE demands. RIS has been shown to be a promising technology to improve the performance of various interference-limited and/or interference-free systems \cite{huang2019reconfigurable, wu2019intelligent, pan2020multicell, 
soleymani2022improper, weinberger2022synergistic, jiang2022interference, santamaria2023icassp}.
A regular passive RIS can provide an additional degree of freedom in the system design by controlling the reflections, which can be used to improve the coverage and/or neutralize interference. To employ a regular RIS, both the transmitter and receiver should be in the reflection space of the regular RIS, which may restrict its applicability. To address this issue, simultaneously-reflect-and-transmit (STAR-) RISs can be employed, which provide a $360^{\circ}$ coverage \cite{9774942, liu2022simultaneously, mu2021simultaneously, xu2021star, soleymani2023energy, soleymani2023spectral}.  In a STAR-RIS, each component can not only transmit, but also can reflect at the same time. Thus, STAR-RIS can cover a wider area and is expected to improve the system performance especially when a regular RIS cannot cover all the users. 

In this paper, we propose NOMA-based IGS schemes to maximize the minimum-weighted rate of users in a MIMO STAR-RIS-assisted BC with IQI. To this end, we extend the proposed scheme in \cite{soleymani2022noma} to MIMO systems with IQI. Note that \cite{soleymani2022noma} proposed schemes to enhance the SE and EE of a multi-cell MISO RIS-assisted BC by considering regular RISs and perfect devices. 
However, in this work, we study STAR-RIS and focus on different modes of STAR-RIS with different assumptions regarding the feasibility sets for optimizing the RIS components. 
We consider two operational modes for STAR-RIS. First, we assume that all the STAR-RIS components can simultaneously operate in both the transmission and reflections modes, which is known as energy splitting (ES). Second, we divide the STAR-RIS components into two groups and operate the components in a group only in the transmission mode, while the components for the other group operate only in the reflection mode. This operational mode is known as the mode switching (MS) scheme. 

Our results show that the NOMA-based IGS scheme can significantly outperform the NOMA-based proper Gaussian signaling (PGS) scheme and/or the IGS scheme with treating interference as noise (TIN).   Moreover, we show that  STAR-RIS can considerably increase the minimum rate of users, even with a relatively low number of STAR-RIS components per users. Furthermore, we show that the MS scheme performs close to the ES schemes with different feasibility sets for STAR-RIS components. MS schemes have lower computational and implementation complexities than the ES schemes, which may make them more practical if their performance is close to ES schemes.   
Additionally, we show that IQI can  reduce the minimum rate of users even if it is compensated by IGS, and neglecting IQI in the design can cause a significant performance loss.

\section{System model}\label{sec-ii}
\begin{figure}[t!]
    \centering
\includegraphics[width=.45\textwidth]{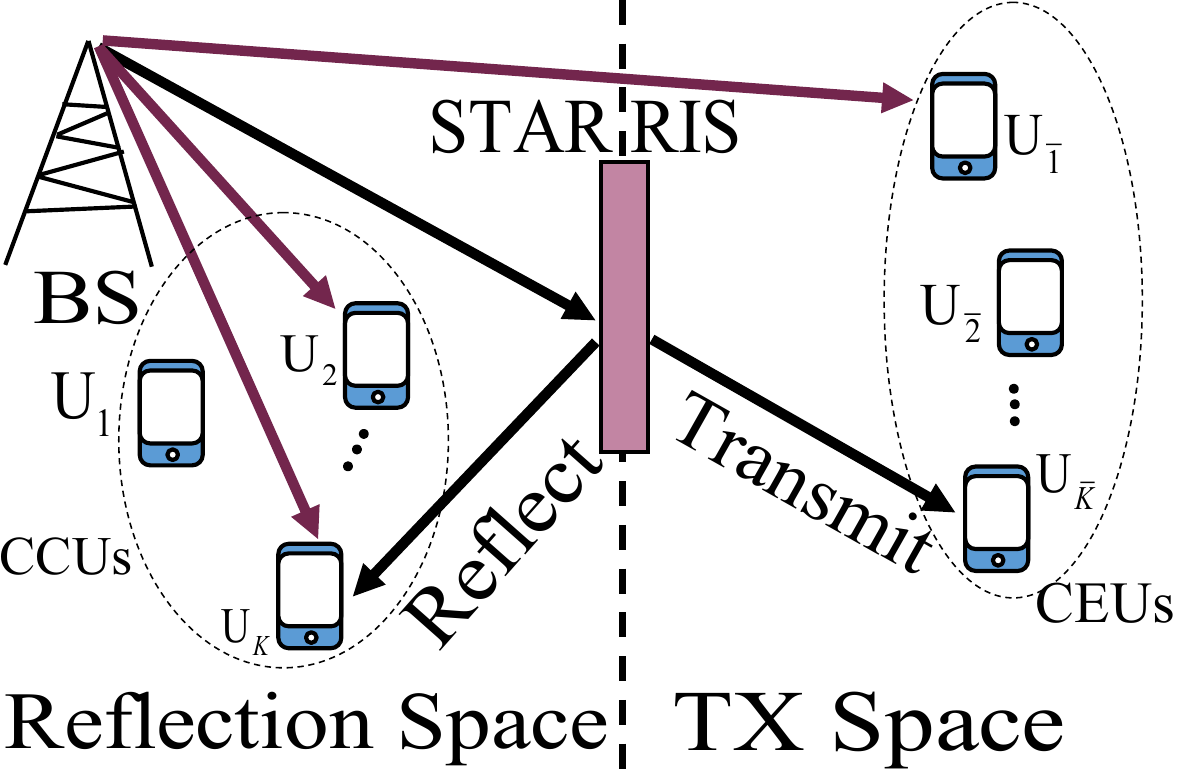}
     \caption{A broadcast channel assisted by a STAR-RIS.}
	\label{Fig-sys-model}
\end{figure}
We consider a single-cell MIMO STAR-RIS-assisted BC, consisting of a BS with $N_{BS}$ antennas and $2K$ users with $N_u$ antennas  each, as shown in Fig. \ref{Fig-sys-model}. We assume that there is a passive STAR-RIS that assists the BS to serve the users.
We further assume that the transceivers suffer from IQI based on the model in \cite{javed2019multiple}. 
We consider a system in which users are located into two clusters with noticeably different distances to the BS, similar to \cite{tuan2019non}. Indeed, we assume that
there are two types of users in each cell: cell-centric  users (CCUs) and cell-edge  users (CEUs), where each cluster has $K$ users. 
We randomly divide the users into $K$ pairs (one CCU and one CEU), as in \cite{tuan2019non}, each pair being served by NOMA. 
Note that the proposed scheme in this paper can be applied to other scenarios with a given user pair/ordering scheme. In this work, our main focus is on optimizing the transmission parameters and STAR-RIS components and leave obtaining the optimal user pairing/ordering for future work.

\subsection{STAR-RIS model}\label{sec-ris}
A STAR-RIS can provide a $360^{\circ}$ coverage since it can simultaneously transmit and reflect. In a STAR-RIS-assisted system, there can be two spaces for each STAR-RIS: transmission and reflection, as depicted in Fig. 
\ref{Fig-sys-model}. In the reflection space, users receive signals through a reflection from STAR-RIS, while in the transmission space, the STAR-RIS transmits signals to users. Thus, each user belongs to either a transmission space or a reflection space of a STAR-RIS. Hence, the channel between the BS and user $k$ is \cite[Eq. (2)]{mu2021simultaneously} 
\begin{equation}\label{ch}
\mathbf{H}_{k}\!\left(\{\bm{\Theta}^r,\bm{\Theta}^t\}\right)\!
=
{\mathbf{G}_{k}\bm{\Theta}^{r/t}\mathbf{G}}
+\!
{\mathbf{F}_{k}}
\in \mathbb{C}^{N_u\times N_{BS}}
,
\end{equation}
where $\mathbf{F}_{k}$ 
is the direct link between the BS and user $k$, $\mathbf{G}$ 
is the channel matrix between the BS and the STAR-RIS, 
$\mathbf{G}_{k}$ 
is the channel vector between the STAR-RIS and user $k$, 
$\bm{\Theta}^r
=\text{diag}\left(\theta_{1}^r, \theta_{2}^r,\cdots,\theta_{{N_{RIS}}}^r\right)$, 
is a diagonal matrix, containing all the reflection coefficients of the STAR-RIS and 
$\bm{\Theta}^t
=\text{diag}\left(\theta_{1}^t, \theta_{2}^t,\cdots,\theta_{{N_{RIS}}}^t\right)$ 
is a diagonal matrix, containing all the transmission coefficients of the STAR-RIS. If user $k$ is in the reflection (or transmission) space of the STAR-RIS, its channel can be optimized by $\bm{\Theta}^r$ (or $\bm{\Theta}^t$). Hereafter, for notational simplicity, we drop the dependency of the channels on the RIS components and represent channels as $\mathbf{H}_{k}$ for all $k$.

There are three different assumptions for modeling the relation between the reflection and transmission coefficients of each STAR-RIS component. First, the amplitudes of the reflection and transmission coefficients are related as \cite[Eq. (2)]{xu2021star}
\begin{equation}\label{1}
|\theta_{i}^{r}|^2+|\theta_{i}^{t}|^2\leq 1, \forall i,
\end{equation}
which is a convex constraint. We call the feasibility sets for this model $\mathcal{T}_{U}$, given by \cite[Eq. (9)]{soleymani2023spectral}. Second, the amplitudes of the reflection and transmission coefficients are related as \cite[Eq. (1)]{liu2021star}: $|\theta_{i}^{r}|^2+|\theta_{i}^{t}|^2= 1$ for all $i$. This non-convex constraint can be written as the two constraints:  \eqref{1} and
\begin{equation}\label{2}
|\theta_{i}^{r}|^2+|\theta_{i}^{t}|^2\geq 1, \forall i.
\end{equation}
Indeed, in this model, there is the additional constraint \eqref{2}, comparing to $\mathcal{T}_{U}$, which implies that the second model is more stringent. We call the feasibility set of the second model $\mathcal{T}_{I}$, given by \cite[Eq. (10)]{soleymani2023spectral}. Finally, in the third model, the amplitudes of the reflection and transmission coefficients are related to as in $\mathcal{T}_{I}$, and the phases are related to as $\angle \theta_{i}^{r}=\angle \theta_{i}^{r}\pm \pi/2$ for all $i$ \cite{9774942, liu2022simultaneously}.
The phase relation can be also modeled as the following convex constraint \cite[Lemma 1]{soleymani2023spectral}
\begin{equation}\label{3}
|\theta_{i}^{r}\pm\theta_{i}^{t}|^2\leq 1, \forall i.
\end{equation}
We represent the feasibility set for the third model by $\mathcal{T}_{N}$, given by \cite[Eq. (11)]{soleymani2023spectral}. It is evident that $\mathcal{T}_{N}\subset\mathcal{T}_{I}\subset\mathcal{T}_{U}$. Thus, $\mathcal{T}_{U}$ should perform not worse than $\mathcal{T}_{I}$, which performs not worse than $\mathcal{T}_{N}$. The set $\mathcal{T}_{U}$ is convex; however, $\mathcal{T}_{N}$ and $\mathcal{T}_{I}$ are non-convex because of the constraint \eqref{2}.

\subsection{Signal model}
The BS aims at transmitting the signal  
\begin{equation}
\mathbf{x}=\sum_{k=1}^{2K}\mathbf{x}_{k}
\in\mathbb{C}^{N_{BS}\times 1},
\end{equation}
 where $\mathbf{x}_{k}$ is the signal intended for user $k$. 
However, due to IQI,   the actual transmitted signal at the BS is 
a WLT of $\mathbf{x}$ as ${\bf x}^t=\mathbf{\Gamma}_{1,t}{\bf x}+\mathbf{\Gamma}_{2,t}{\bf x}^*$, where $\mathbf{\Gamma}_{1,t}=\frac{\mathbf{I}+\mathbf{A}_Te^{j \bm{\phi}_T}}{2}$ and $\mathbf{\Gamma}_{2,t}={\bf I}-\mathbf{\Gamma}_{1,t}^*$  \cite{soleymani2020improper}. 
The matrices $\mathbf{A}_T$ and $\bm{\phi}_T$ are diagonal matrices that correspond to, respectively, the amplitude and phase errors between the two quadrature signal paths at the transmitter side \cite{javed2019multiple}. If $\mathbf{A}_T={\bf I}$ and $\bm{\phi}_T={\bf 0}$, there is no IQI. Otherwise, ${\bf x}^t$ can be improper even if $\mathbf{x}$ is proper.

The received signal at user $k$ is
${\bf y}_{k}
=
{\mathbf{H}_{k}\mathbf{x}^t
}
+
{{\bf n}_{k}}
$,
where 
 ${\bf n}_{k}$ is zero-mean   additive  white Gaussian noise with variance $\sigma^2{\bf I}$. 
 Due to IQI at  the receiver of user $k$, the output signal is a widely linear transformation of ${\bf y}_{k}$ as ${\bf y}^r_k=\mathbf{\Gamma}_{1,r}{\bf y}_k+\mathbf{\Gamma}_{2,r}{\bf y}^*_k$, where $\mathbf{\Gamma}_{1,r}=\frac{\mathbf{I}+\mathbf{A}_Re^{j \bm{\phi}_R}}{2}$ and $\mathbf{\Gamma}_{2,r}={\bf I}-\mathbf{\Gamma}_{1,r}^*$  \cite{soleymani2020improper}. 
Similarly, the matrices $\mathbf{A}_R$ and $\bm{\phi}_R$ are diagonal matrices, corresponding to the amplitude and phase errors of each branch at the receiver side, respectively \cite{javed2019multiple}.

We employ the real-decomposition method for modeling and optimizing improper signals, similar to \cite{soleymani2020improper, soleymani2022improper}. The output signal at the receiver side, using the real-decomposition method, is
\begin{align}\label{rec-sig=2}
\underline{\mathbf{y}}_{k}
=
{\underline{\mathbf{H}}_{k}
\sum_{i=1}^{2K}\underline{\mathbf{x}}_i
}
+
\underline{\mathbf{n}}_{k},
\end{align}
 where $\underline{\mathbf{y}}_{k}=\left[\begin{array}{cc}\mathfrak{R}\{{\bf y}_{k}\}^T&\mathfrak{I}\{{\bf y}_{k}\}^T\end{array}\right]^T$, and
 $\underline{\mathbf{x}}_k=\left[\begin{array}{cc}\mathfrak{R}\{{\bf x}_k\}^T&\mathfrak{I}\{{\bf x}_k\}^T\end{array}\right]^T$ for all $k$. 
 Moreover,  
$\underline{\mathbf{n}}_{k}$ is the effective noise at the output of user $k$, and $\underline{\mathbf{H}}_{k}$ is the effective channel in the real domain, given by \cite[Lemma 1]{soleymani2022rate}.
The signals $\underline{\mathbf{x}}_k$ and $\underline{\mathbf{n}}_k$ are zero-mean. We represent the covariance matrices of $\underline{\mathbf{x}}_k$ and $\underline{\mathbf{n}}_k$ by, respectively, ${\bf P}_k$ and $\underline{\bf C}_n$, where ${\bf P}_k$ is an optimization parameter, while $\underline{\bf C}_n$ is given by \cite[Lemma 1]{soleymani2022rate}.  
The matrix $\underline{\bf C}_n$ is equal to $\frac{\sigma^2}{2}{\bf I}$ if there is no IQI at the receivers. Otherwise, $\underline{\mathbf{n}}_k$ is improper and its covariance matrix is not equal to a scaled identity matrix.

\subsection{NOMA}
We treat the inter-cluster interference as noise and employ IGS and spatial dimensions to handle it. However, we employ NOMA to manage intra-cluster interference.
That is, we assume that the CCU $k$ first decodes the message of its paired CEU $\bar{k}=K+k$ and cancels the message from the received signal. Thus, the achievable rate of user $k$ is  
\begin{align}
{r}_{{k}}&=\frac{1}{2}\log_2\left|{\bf I}+
{\bf D}^{-1}_{{k}}
{\bf S}_{{k}}
 \right|
=
\underbrace{
\frac{1}{2}\log_2\left|{\bf D}_{{k}}
+
{\bf S}_{{k}}
\right|
}_{{r}_{{k},1}}
-
\underbrace{
\frac{1}{2}\log_2\left|{\bf D}_{{k}}
\right|
}_{\bar{r}_{{k},2}},
\end{align}
where $\mathbf{D}_{k}
=
\sum_{j= 1,j\neq k,\bar{k}}^{2K}
\underline{\mathbf{H}}_{k}
\mathbf{P}_{j}
\underline{\mathbf{H}}_{k}^T
+
\underline{\bf C}_n
$, and $\mathbf{S}_{k}=\underline{\mathbf{H}}_{k}
\mathbf{P}_{k}
\underline{\mathbf{H}}_{k}^T$.
On the contrary, the CEU $\bar{k}$ treats all the received signals as noise while decoding its own signal. The message for user $\bar{k}$ should be transmitted at a rate that is decodable for both users $k$ and $\bar{k}$.
Thus, the maximum transmission rate for user $\bar{k}$ is 
\begin{equation}\label{eq-12}
r_{\bar{k}}=\min\left(\bar{r}_{\bar{k}},\bar{r}_{k\rightarrow\bar{k}}\right),
\end{equation}
where $\bar{r}_{\bar{k}}$ is the maximum decoding rate of ${\bf x}_{\bar{k}}$ at  user ${\bar{k}}$ treating all other signals as noise, which is given by 
\begin{align}
\bar{r}_{\bar{k}}&=\frac{1}{2}\log_2\left|{\bf I}+
{\bf D}^{-1}_{\bar{k}}
{\bf S}_{\bar{k}}
 \right|
=
\underbrace{
\frac{1}{2}\log_2\left|{\bf D}_{\bar{k}}
+
{\bf S}_{\bar{k}}
\right|
}_{\bar{r}_{\bar{k},1}}
-
\underbrace{
\frac{1}{2}\log_2\left|{\bf D}_{\bar{k}}
\right|
}_{\bar{r}_{\bar{k},2}},
\label{eq-22-2}
\end{align}
where 
$\mathbf{D}_{l\bar{k}}
=
\sum_{j= 1,j\neq \bar{k}}^{2K}
\underline{\mathbf{H}}_{\bar{k}}
\mathbf{P}_{j}\underline{\mathbf{H}}_{\bar{k}}^T
+
\underline{\bf C}_n
$, and $\mathbf{S}_{\bar{k}}=\underline{\mathbf{H}}_{\bar{k}}
\mathbf{P}_{\bar{k}}
\underline{\mathbf{H}}_{\bar{k}}^T$.
Moreover, $\bar{r}_{k\rightarrow\bar{k}}$ is the maximum decoding rate of ${\bf x}_{\bar{k}}$ at  user $k$, while treating all other signals as noise, given by
\begin{align}
\nonumber
\bar{r}_{k\rightarrow\bar{k}}&=
\underbrace{
\frac{1}{2}\log_2\left|{\bf D}_{k}
+
{\bf S}_{k}
+
{\bf S}_{k\rightarrow\bar{k}}
\right|
}_{\bar{r}_{k\rightarrow\bar{k},1}
}
-
\underbrace{
\frac{1}{2}\log_2\left|{\bf D}_{k}
+
{\bf S}_{k} 
\right|
}_{\bar{r}_{lk\rightarrow\bar{k},2}
}
\\
&
=\frac{1}{2}\log_2\left|{\bf I}+
\left({\bf D}_{k}
+
{\bf S}_{k}
%
\right)^{-1}
 {\bf S}_{k\rightarrow\bar{k}}
\right|,
\label{eq-26}
\end{align}
where ${\bf S}_{k\rightarrow\bar{k}}=
\underline{{\bf H}}_{k}
{\bf P}_{\bar{k}}\underline{{\bf H}}_{k}^T
$.

\subsection{Problem statement}
In this paper, we aim at maximizing the minimum weighted rate, which can be written as 
\begin{align}\label{opt}
\underset{
r,\bm{\Theta}\in\mathcal{T}, \{\mathbf{P}\}\in\mathcal{P}}
{\max} \hspace{0.2cm}&r,
&
\text{s.t.}\hspace{0.4cm}& \lambda_{k}r_{k}\geq r, & k=1,\cdots,2K,
\end{align}
where $\lambda_{k}$ is the corresponding weight for user $k$, $\{\mathbf{P}\}$ is the set of covariance matrices, and $\mathcal{P}$ is the feasibility set of the covariance matrices, which is given by \cite[Eq. (23)]{soleymani2022noma} for IGS and by \cite[Eq. (24)]{soleymani2022noma} for PGS schemes.

\section{Proposed scheme}
To solve \eqref{opt}, we employ alternating optimization (AO) and  an MM-based technique. 
That is, we first fix $\{\bm{\Theta}^{(t-1)}\}$, and optimize over covariance matrices to obtain $\{{\bf P}^{(t)}\}$. We then alternate and optimize over RIS components for fixed covariance matrices $\{{\bf P}^{(t)}\}$ to find $\{\bm{\Theta}^{(t)}\}$. We iterate this procedure until a convergence criterion is met.

\subsection{Optimizing transmit covariance matrices}
Solving \eqref{opt} for fixed $\{\bm{\Theta}^{(t-1)}\}$ is very similar to the scheme in \cite[Sec. III.A]{soleymani2022noma}. For the sake of completeness, we briefly describe the scheme and refer the reader to \cite{soleymani2022noma} for more details.
The optimization problem \eqref{opt} for fixed $\{\bm{\Theta}^{(t-1)}\}$ is non-convex. Thus, we employ an MM-based algorithm to find a solution for  this problem. That is, we first find suitable surrogate functions for the rates. To this end, we employ 
\cite[Lemma 3]{soleymani2022rate} to approximate the convex part of the rates by an affine function, which results in the following concave lower bounds
\begin{align}
\nonumber
 r_{k}&\geq \tilde{r}_{k}=r_{k,1}
-r_{k,2}^{(t-1)}
\\
\label{l-r-lk}
&-\!\!\!\!\sum_{j=1,j\neq k,\bar{k}}^{2K}\!\!\!\!
\text{Tr}\!\left(\!
\frac{
\underline{\mathbf{H}}_{k}^T
(\mathbf{D}_{k}^{(t-1)})^{-1}
\underline{\mathbf{H}}_{k}
}
{2\ln 2}\!
\left(\mathbf{P}_{j}-\mathbf{P}_{j}^{(t-1)}\right)\!\!\!
\right)\!,
\\
\nonumber
\tilde{r}_{\bar{k}}^l&=\bar{r}_{\bar{k},1}
-\bar{r}_{\bar{k},2}^{(t-1)}
\\&
\label{l-r-lk-bar}
-\!\!\!\!
\sum_{j=1,j\neq k}^{2K}\!\!\!
\text{Tr}\!
\left(
\frac{
\underline{\mathbf{H}}_{\bar{k}}^T
(\mathbf{D}_{\bar{k}}^{(t-1)})^{-1}
\underline{\mathbf{H}}_{\bar{k}}
}
{2\ln 2}
\left(\mathbf{P}_{j}-\mathbf{P}_{j}^{(t-1)}\right)\!\!
\right),
\\
\nonumber
\tilde{r}_{k\rightarrow\bar{k}}^l&=\bar{r}_{k\rightarrow\bar{k},1}
-\bar{r}_{k\rightarrow\bar{k},2}^{(t-1)}
-\!\!\!\sum_{j=1,j\neq\bar{k}}^{2K}
\!\!\!
\text{Tr}\!
\left(\!\!
\frac{
\underline{\mathbf{H}}_{k}^T
}
{2\ln 2}\right.
\\
&
\left.
\times\left(
\mathbf{D}_{k}^{(t-1)}\!\!+\!\mathbf{S}_{k}^{(t-1)}
\right)^{-1}
\underline{\mathbf{H}}_{k}\left(\mathbf{P}_{ij}-\mathbf{P}_{ij}^{(t-1)}\right)
\right),
\label{l-r-lk-bar-2}
\end{align}
where $r_{k,2}^{(t-1)}$, $r_{\bar{k},2}^{(t-1)}$, $r_{k\rightarrow\bar{k},2}^{(t-1)},$ $\mathbf{S}_{k}^{(t-1)}$,  $\mathbf{D}_{k}^{(t-1)}$, and $\mathbf{D}_{\bar{k}}^{(t-1)}$ are the corresponding initial values at this optimization step. 
Note that the concave lower bound for the rate of the CEU $\bar{k}$ is 
$\tilde{r}_{\bar{k}}=\min\left(\tilde{r}_{\bar{k}}^l,\tilde{r}_{k\rightarrow\bar{k}}^l\right).$
 Substituting the concave lower bounds for the rates in the optimization problem \eqref{opt} for fixed $\{\bm{\Theta}^{(t-1)}\}$ results in a convex optimization problem, which can be efficiently solved by existing numerical tools.

\subsection{Optimizing RIS components}
In this subsection, we update the STAR-RIS coefficients by solving \eqref{opt} for fixed $\{{\bf P}^{(t)}\}$. 
Unfortunately, \eqref{opt} for fixed $\{{\bf P}^{(t)}\}$ is not convex since the rates are not concave in $ \{\bm{\Theta}\}$. Moreover, the feasibility set for STAR-RIS components is not convex for the sets $\mathcal{T}_I$ and $\mathcal{T}_N$. To update $ \{\bm{\Theta}\}$, we propose a suboptimal scheme based on MM. In the following, we first consider the ES scheme.  Then we mention how the scheme can be modified to be applied to MS schemes. 

To solve \eqref{opt} for fixed $\{{\bf P}^{(t)}\}$, we should first obtain suitable concave lower bounds for the rates. To this end, we employ the bounds in \cite[Lemma 4]{soleymani2022rate}, which results in the following lower bounds
\begin{align}
\nonumber
\hat{r}_{k}&=r_{k}^{(t-1)}
-\frac{1}{2\ln 2}\text{{ Tr}}\left(
\bar{\mathbf{S}}_{k}\bar{\mathbf{D}}_{k}^{-1}
\right)
+
\frac{1}{\ln 2}
\text{{ Tr}}\left(
{\mathbf{A}_{k}}
\underline{{\bf H}}_{k}
\right)
\\
&\hspace{.1cm}-
\frac{1}{2\ln 2}\!
\text{{ Tr}}\!
\left(
(\bar{\mathbf{D}}^{-1}_{k}-(\bar{\mathbf{S}}_{k}+ \bar{\mathbf{D}}_{k})^{-1})
(\mathbf{D}_{k}+\mathbf{S}_{k})
\right)\!
,
\\
\nonumber
\hat{r}_{\bar{k}}^l&=r_{\bar{k}}^{(t-1)}
-\frac{1}{2\ln 2}\text{{ Tr}}\left(
\bar{\mathbf{S}}_{\bar{k}}\bar{\mathbf{D}}_{\bar{k}}^{-1}
\right)
+
\frac{1}{\ln 2}
\text{{ Tr}}\left(
{\mathbf{A}_{\bar{k}}}
\underline{{\bf H}}_{\bar{k}}
\right)
\\
&\hspace{.1cm}-
\frac{1}{2\ln 2}\!
\text{{ Tr}}\!
\left(
(\bar{\mathbf{D}}^{-1}_{\bar{k}}-(\bar{\mathbf{S}}_{\bar{k}}+ \bar{\mathbf{D}}_{\bar{k}})^{-1})
(\mathbf{D}_{\bar{k}}+\mathbf{S}_{\bar{k}})
\right)
\!
,
\\
\nonumber
\hat{r}_{k\rightarrow\bar{k}}^l&=r_{k\rightarrow\bar{k}}^{(t-1)}
-\frac{1}{2\ln 2}\left[
\text{{ Tr}}\left(
\bar{{\bf S}}_{k\rightarrow\bar{k}}(\bar{\mathbf{D}}_{{k}}+\bar{\mathbf{S}}_{{k}})^{-1}
\right)
\right.
\\
&\nonumber
\hspace{.2cm}
+
2
\text{{ Tr}}\left(
{\mathbf{A}_{k\rightarrow\bar{k}}}
\underline{{\bf H}}_{\bar{k}}
\right)
-
\text{{ Tr}}\left(
((\bar{\mathbf{D}}_{{k}}+\bar{\mathbf{S}}_{{k}})^{-1}
\right.
\\
&\hspace{.2cm}
\left.\left.
-({\bf S}_{k\rightarrow\bar{k}}\!
+\bar{\mathbf{S}}_{{k}}\!
+ \bar{\mathbf{D}}_{{k}})^{-1})
(\mathbf{D}_{{k}}\!+\mathbf{S}_{{k}}\!+{\bf S}_{k\rightarrow\bar{k}})
\right)
\right],
\end{align}
where $\bar{\mathbf{S}}_{k}$, $\bar{\mathbf{S}}_{\bar{k}}$, $\bar{\mathbf{S}}_{k\rightarrow\bar{k}}$, 
$\bar{\mathbf{D}}_{k}$, and $\bar{\mathbf{D}}_{\bar{k}}$ are the corresponding initial values at this optimization step. 
Moreover,
$\mathbf{A}_{k}={\bf P}_{k}\underline{\mathbf{H}}_{k}^T(\{\bm{\Theta}^{(t-1)}\}) \bar{\mathbf{D}}_{k}^{-1}$, 
$\mathbf{A}_{\bar{k}}={\bf P}_{\bar{k}}\underline{\mathbf{H}}_{\bar{k}}^T(\{\bm{\Theta}^{(t-1)}\}) \bar{\mathbf{D}}_{\bar{k}}^{-1}$, and 
$\mathbf{A}_{k\rightarrow\bar{k}}={\bf P}_{\bar{k}}\underline{\mathbf{H}}_{k}^T(\{\bm{\Theta}^{(t-1)}\}) (\bar{\mathbf{D}}_{k}+\bar{\mathbf{S}}_{k})^{-1}$. 
Finally, the concave lower bound for the CEU $\bar{k}$ is $\hat{r}_{\bar{k}}=\min\left( \hat{r}_{\bar{k}}^l,\hat{r}_{k\rightarrow\bar{k}}^l\right)$. 
Substituting the concave lower bounds for the rates in the problem \eqref{opt}, we have 
\begin{align}\label{opt2}
\underset{
r,\bm{\Theta}\in\mathcal{T}
}
{\max} \hspace{0.2cm}&r,
&
\text{s.t.}\hspace{0.4cm}& \lambda_{k}\hat{r}_{k}\geq r, & k=1,\cdots,2K,
\end{align}
which is convex only for $\mathcal{T}_U$. However, since the sets  $\mathcal{T}_I$ and $\mathcal{T}_N$ are not convex sets (because of \eqref{3}), the problem \eqref{opt2} is not convex for these sets.
To convexify \eqref{opt2} for $\mathcal{T}_I$ and $\mathcal{T}_N$, we can apply CCP to the constraint in \eqref{2} since $|\theta_{i}^{r}|^2+|\theta_{i}^{t}|^2$ for all $i$ is a convex function and should be converted to a concave function. Additionally, we can relax \eqref{3} for a faster convergence. 
Thus,  \eqref{2} can be approximated as
\begin{multline}\label{eq-50-6}
|\theta_{i}^{r^{(t-1)}}|^2+2\mathfrak{R}\left(\theta_{i}^{r^{(t-1)}}(\theta_{i}^r-\theta_{i}^{r^{(t-1)}})^*\right)
+\!
|\theta_{i}^{t^{(t-1)}}|^2\!
\\
+\!
2\mathfrak{R}\!\left(
\!
\theta_{i}^{t^{(t-1)}}(\theta_{i}^t-\theta_{i}^{t^{(t-1)}})^*
\!
\right)
\!\!
\geq 
\!
1-\epsilon,\forall i,
\end{multline}
where $\epsilon>0$.
The constraint \eqref{eq-50-6} is convex since 
it is linear in $\theta_{i}^r$ and $\theta_{i}^t$.
Therefore, \eqref{opt2} for $\mathcal{T}_I$ can be approximated as 
\begin{subequations}\label{opt-t-sur2}
\begin{align}
\underset{
r,\bm{\Theta}
}
{\max} \hspace{0.2cm}&r,
&
\text{s.t.}\hspace{0.4cm}& \lambda_{k}\hat{r}_{k}\geq r, & k=1,\cdots,2K,
\\
&&& \eqref{1},\eqref{eq-50-6}.
\end{align}
\end{subequations}
Moreover, \eqref{opt2} for $\mathcal{T}_N$ can be approximated as 
\begin{subequations}\label{opt-t-sur3}
\begin{align}
\underset{
r,\bm{\Theta}
}
{\max} \hspace{0.2cm}&r,
&
\text{s.t.}\hspace{0.4cm}& \lambda_{k}\hat{r}_{k}\geq r, & k=1,\cdots,2K,
\\
&&& \eqref{1},\eqref{3},\eqref{eq-50-6}.
\end{align}
\end{subequations}
The problems \eqref{opt-t-sur2} and \eqref{opt-t-sur3} are convex and can be efficiently solved. However, their solutions may not satisfy the constraint $|\theta_{i}^{r}|^2+|\theta_{i}^{t}|^2= 1$ due to the relaxation in \eqref{eq-50-6}. 
Thus, to ensure a valid solution, we should normalize the solutions of \eqref{opt-t-sur2} (or \eqref{opt-t-sur3}) as 
\begin{align}\label{eq3300}
\hat{\theta}_{i}^t\!&\!=\!\frac
{{\theta}_{i}^{t^{(\star)}}}
{\sqrt{|{\theta}_{i}^{t^{(\star)}}|^2+|{\theta}_{i}^{r^{(\star)}}|^2}},&\!\!
\hat{\theta}_{i}^r\!&\!=\!\frac
{{\theta}_{i}^{r^{(\star)}}}
{\sqrt{|{\theta}_{i}^{t^{(\star)}}|^2+|{\theta}_{i}^{r^{(\star)}}|^2}},
\end{align}
for all $i$. 
Although $\hat{\theta}_{i}^t$ and $\hat{\theta}_{i}^r$ are feasible, they may not result in a non-decreasing sequence of the minimum rate of users. To address this issue, 
 we update $\bm{\Theta}^r$ and $\bm{\Theta}^t$ such that 
\begin{equation}\label{eq-42-star}
\{\!\bm{\Theta}^{(t)}\!\}\!=\!\!
\left\{\!\!\!\!\!
\begin{array}{lcl}
\{\hat{\bm{\Theta}}\}\!\!&\!\!\!\!\!\!\text{if}\!\!&
\min_{k}\left\{\lambda_{k}r_k\!\!\left(\{\hat{\bm{\Theta}}\}\!\right)\right\}\!\!\geq\!\!\!\!\!
\\
&&
\min_{k}\left\{\lambda_{k}r_k\!\!\left(\{\bm{\Theta}^{(t-1)}\}\!\right)\right\}
\\
\{\bm{\Theta}^{(t-1)}\}&&\text{Otherwise},
\end{array}
\right.
\end{equation}
where $\{\hat{\bm{\Theta}}\}=\{\hat{\bm{\Theta}}^{r},\hat{\bm{\Theta}}^{t}\}$ is given by \eqref{eq3300}. 
This updating step ensures the convergence of the proposed scheme since the framework generates a non-decreasing sequence of the minimum rates of the users.

Now we consider the MS scheme in which each STAR-RIS component either reflects or transmits at a time. In this case, we set ${\theta}_{i}^t=0$ (or ${\theta}_{i}^r=0$) for STAR-RIS components that operate in the reflection (or transmission) mode. Then the other steps remain the same. Specifically, to obtain new $\{\bm{\Theta}\}$, we should solve \eqref{opt2} for $\mathcal{T}_U$, \eqref{opt-t-sur2} for $\mathcal{T}_I$, and \eqref{opt-t-sur3} for $\mathcal{T}_N$. Then we should normalize the solutions for the feasibility sets $\mathcal{T}_I$ and $\mathcal{T}_N$ according to \eqref{eq3300} and update $\{\bm{\Theta}\}$ based on \eqref{eq-42-star}.
Note that the feasibility sets $\mathcal{T}_I$ and $\mathcal{T}_N$ are equivalent for the MS scheme since each STAR-RIS component either transmits or reflects in which the phase relation does not have any impact.

\section{Numerical Results}
In this section, we present some numerical results.
To this end, we consider a scenario in which the CCUs are in the reflection space of the RIS, while the CEUs are in the transmission space of the RIS. Thus, a regular RIS can assist only the CCUs and cannot cover the CEUs. 
However, a STAR-RIS can cover all the users due to its $360^{\circ}$ coverage. 
In order to have a fair comparison, we assume that the number of RIS components for both the regular and STAR-RISs are the same. To better investigate the role of IGS, we divide the numerical results into two subsection. In the first subsection, we investigate the role of IGS as an interference-management technique. To this end, we consider perfect devices since IQI can increase the benefits of IGS, as shown in \cite{soleymani2020improper, javed2019multiple}. Then, we consider the impact of IQI and the importance of HWI-aware techniques in Section IV.B.

\subsection{Impact of interference on spectral efficiency}
\begin{figure}[t!]
    \centering
       \includegraphics[width=.45\textwidth]{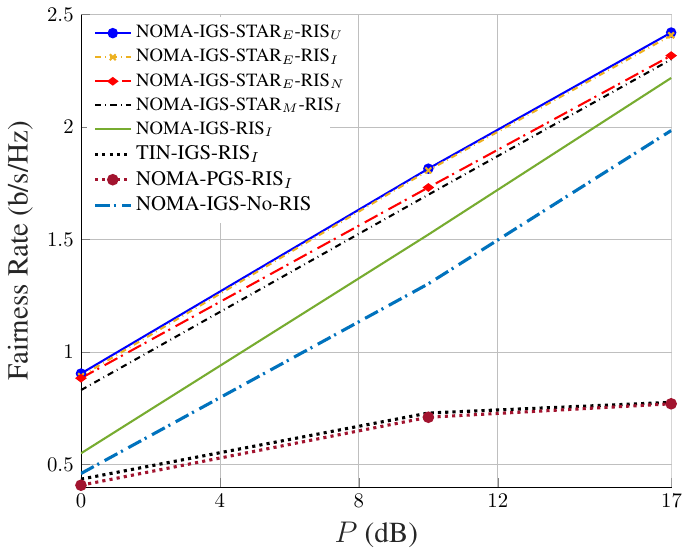}
    \caption{The average fairness rate versus $P$ for $N_{BS}=N_u=2$, $N_{RIS}=60$, and $K=3$.}
	\label{Fig-sr} 
\end{figure}
Fig. \ref{Fig-sr}  shows the average fairness rate versus $P$ for $N_{BS}=N_u=2$,  $N_{RIS}=60$, and $K=3$. As can be observed, the proposed NOMA-based IGS scheme with ES and the feasibility set $\mathcal{T}_U$ outperforms the other schemes, while the  NOMA-based IGS scheme with ES and $\mathcal{T}_I$ performs very close to the ES scheme with $\mathcal{T}_U$. 
Moreover, we observe that the MS scheme performs very close to the ES schemes, and the performance gap between the MS scheme and the ES scheme with the feasibility set $\mathcal{T}_N$ is minor, especially at higher SNR regimes. 
Note that the computational and/or implementation complexities of MS schemes are less than ES schemes, which can compensate the small performance gap, comparing to the ES schemes. 

In Fig. \ref{Fig-sr}, we can also observe that IGS and/or NOMA can highly improve the system performance. Interestingly, the NOMA-based IGS scheme without RIS can significantly outperform the NOMA-based PGS scheme as well as the IGS scheme with TIN for RIS-assisted systems. This result, indeed, shows the importance of interference-management techniques in highly overloaded STAR-RIS-assisted systems. Moreover, it shows that RIS alone cannot handle interference in overloaded systems, and we have to employ powerful interference-management techniques to fully reap (STAR-)RIS benefits.

\begin{figure}[t!]
    \centering
       \includegraphics[width=.45\textwidth]{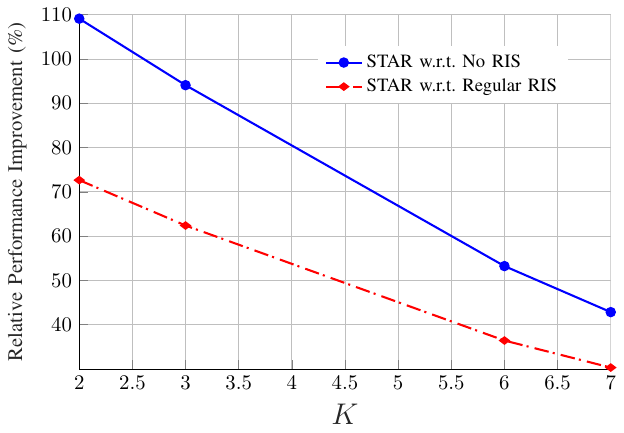}
    \caption{The average performance improvement by employing STAR-RIS versus $K$ for $N_{BS}=N_u=2$, and $N_{RIS}=60$.}
	\label{Fig-ris} 
\end{figure}
Fig. \ref{Fig-ris} shows the benefits of employing STAR-RIS versus $K$ for $N_{BS}=N_u=2$, and $N_{RIS}=60$. As can be observed,  STAR-RIS can significantly improve the system performance even over a regular RIS. For a fixed number of STAR-RIS components, the benefits of STAR-RIS decrease with the number of users. However, the benefits are still significant when there are less than five components per users, i.e., when there are 14 users in the network (for $K=7$). Note that the total number of users is $2K$.

\begin{figure}[t!]
    \centering
       \includegraphics[width=.3\textwidth]{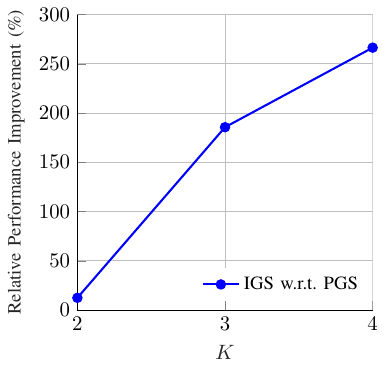}
\caption{IGS with respect to PGS.}
    \caption{The average performance improvement by employing IGS versus $K$ for $N_{BS}=N_u=2$, and $N_{RIS}=60$.}
	\label{Fig-igs} 
\end{figure}
Fig. \ref{Fig-igs} shows the benefits of employing IGS 
versus $K$ for $N_{BS}=N_u=2$, and $N_{RIS}=60$. The IGS benefits are computed by comparing the performance of the NOMA-based IGS scheme with the NOMA-based PGS scheme for RIS-assisted systems.As can be observed, IGS 
can significantly improve the system performance. Furthermore, the   benefits of IGS and/or NOMA increase with the number of users, which is in line with our previous studies in \cite{soleymani2020improper, soleymani2022noma}.
The reason is that, the higher the number users is, the higher interference exists, which makes the use of advanced interference management techniques more necessary and their benefit more significant.
\subsection{Impact of IQI on spectral efficiency}
In this subsection, we consider the impact of IQI on the system performance. We assume that the  IQI parameters at all users are similar. In other words, we set $\mathbf{A}_T=\mathbf{A}_R$ and $\bm{\phi}_T=\bm{\phi}_R$. Similar to \cite{javed2019multiple}, we assume that  $\mathbf{A}_T=a_t{\bf I}$ and $\bm{\phi}_T=\phi{\bf I}$, where $\phi=5^{\circ}$. In the previous subsection, we study the impact of IGS and NOMA as interference-management techniques. Moreover, we investigate the performance of STAR-RIS. thus, in this subsection, we consider only the NOMA-based IGS scheme for STAR-RIS-assisted systems with MS to study the impact of IQI. 

\begin{figure}[t!]
    \centering
       \includegraphics[width=.45\textwidth]{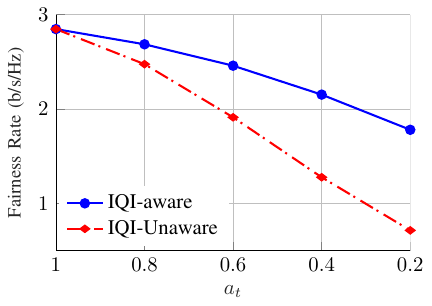}
    \caption{The impact of IQI on the minimum rate of users for $N_{BS}=N_u=2$, $K=2$ and $N_{RIS}=30$.}	\label{Fig-iqi} 
\end{figure}
Fig. \ref{Fig-iqi} shows the impact of IQI on the minimum rate of users for $N_{BS}=N_u=2$, $K=2$ and $N_{RIS}=30$. As can be observed, the minimum rate of users decrease with the IQI level, where there is more than around $35\%$ performance loss for $a_t=0.2$ even when the IQI is considered and compensated in the design.
Such performance loss can even increase up to more than $70\%$ if IQI is overlooked in the system design.

\section{Conclusion}
This paper proposed a scheme to improve the minimum rate of a MIMO STAR-RIS-assisted BC with HWI by employing IGS and NOMA. We considered three feasibility sets for STAR-RIS components and proposed different schemes to optimize STAR-RIS components. We showed that STAR-RIS can significantly improve the minimum rate of the users. Furthermore, we showed that the MS scheme can perform very close to the ES scheme, especially when the phases of the reflection and transmission coefficients are dependent. Additionally, we showed that the NOMA-based IGS scheme can highly outperform the NOMA-based PGS and/or the IGS scheme with TIN, and the benefits of IGS increases with the number of users, which is in line with our previous study in \cite{soleymani2022noma}. Finally, we showed that IQI can highly degrade the system performance even if it is compensated in the system design.

\section*{Acknowledgment}
The work of I. Santamaria was supported by Ministerio de Ciencia e Innovaci\`on and AEI (10.13039/501100011033), under grants PID2019104958RB-C43(ADELE) and PID2022-137099NB-C43 (MADDIE). The work of Eduard Jorswieck was supported in part by the Federal Ministry of Education and Research (BMBF, Germany) as part of the 6G Research and Innovation Cluster 6G-RIC under Grant 16KISK031.

\bibliographystyle{IEEEtran}
\bibliography{ref2}
\end{document}